\documentclass[twocolumn,aps,prl,superscriptaddress,showpacs,secnumroman,
showkeys]{revtex4}
\usepackage{amsmath,bm}%
\usepackage{graphicx}%

\begin{document}

\title{Tracing the Evolution of Temperature in Near Fermi Energy Heavy Ion 
Collisions}  

\author{J. Wang}
\thanks{On leave from the Institue of Modern Physics, Chinese Academy
of Science, Lanzhou 73, China.}
\affiliation{Cyclotron Institute, Texas A\&M University, College Station, 
Texas 77843}
\email[E-mail at:]{wang@comp.tamu.edu}
\author{R. Wada}
\affiliation{Cyclotron Institute, Texas A\&M University, College Station, 
Texas 77843}
\affiliation{Riken, Cyclotron Center, 2-1 Hirosawa, Wako, Saitama, Japan 
351-0198}
\thanks{Riken Collaborative Scientist, Beam Technology Division, 
Cyclotron Center}
\author{T. Keutgen}
\thanks{Now at FNRS and IPN, Universit\'e Catholique de Louvain, B-1348 
Louvain-Neuve, Belgium}
\author{K. Hagel}
\author{Y. G. Ma}
\thanks{on leave from Shanghai Institute of Nuclear Research,
Chinese Academy of Sciences, Shanghai 201800, China}
\author{M. Murray}
\thanks{Now at University of Kansas, Lawrence, Kansas 66045-7582}
\author{L. Qin}
\author{A. Botvina}
\author{S. Kowalski}
\author{T. Materna}
\author{J. B. Natowitz}
\affiliation{Cyclotron Institute, Texas A\&M University, College Station, 
Texas 77843}
\author{R. Alfarro}
\affiliation{Instituto de Fisica, Universidad National Autonoma de Mexico, 
Apactado Postal 20-364 01000, Mexico City, Mexico}
\author{J. Cibor}
\affiliation{Institute of Nuclear Physics, ul. Radzikowskiego 152, PL-31-342 
Krakow, Poland}
\author{M. Cinausero}
\affiliation{INFN, Laboratori Nazionali di Legnaro, I-35020 Legnaro, Italy}
\author{Y. El Masri}
\affiliation{FNRS and IPN, Universit\'e Catholique de Louvain, B-1348 
Louvain-Neuve, Belgium}
\author{D. Fabris}
\affiliation{INFN and Dipartimento di Fisica dell' Universit\'a di Padova, 
I-35131 Padova, Italy}
\author{E. Fioretto}
\affiliation{INFN and Dipartimento di Fisica dell' Universit\'a di Padova, 
I-35131 Padova, Italy}
\author{A. Keksis}
\affiliation{Cyclotron Institute, Texas A\&M University, College Station, 
Texas 77843}
\author{ M. Lunardon}
\affiliation{INFN and Dipartimento di Fisica dell' Universit\'a di Padova, 
I-35131 Padova, Italy}
\author{A. Makeev}
\author{N. Marie}
\thanks{Now at LCP Caen, ISMRA, IN2P3-CNRS, F-14050 Caen, France}
\author{ E. Martin}
\affiliation{Cyclotron Institute, Texas A\&M University, College Station, 
Texas 77843}
\author{Z. Majka}
\affiliation{Jagellonian University, M Smoluchowski Institute of Physics, 
PL-30059, Krakow, Poland}
\author{A. Martinez-Davalos}
\author{A. Menchaca-Rocha}
\affiliation{Instituto de Fisica, Universidad National Autonoma de Mexico, 
Apactado Postal 20-364 01000, Mexico City, Mexico}
\author{G. Nebbia}
\affiliation{INFN and Dipartimento di Fisica dell' Universit\'a di Padova, 
I-35131 Padova, Italy}
\author{G. Prete}
\affiliation{INFN, Laboratori Nazionali di Legnaro, I-35020 Legnaro, Italy}
\author{V. Rizzi}
\affiliation{INFN and Dipartimento di Fisica dell' Universit\'a di Padova, 
I-35131 Padova, Italy}
\author{A. Ruangma}
\author{D. V. Shetty}
\author{G. Souliotis}
\affiliation{Cyclotron Institute, Texas A\&M University, College Station, 
Texas 77843}
\author{P. Staszel}
\affiliation{Jagellonian University, M Smoluchowski Institute of Physics, 
PL-30059, Krakow, Poland}
\author{M. Veselsky}
\affiliation{Cyclotron Institute, Texas A\&M University, College Station, 
Texas 77843}
\author{G. Viesti}
\affiliation{INFN and Dipartimento di Fisica dell' Universit\'a di Padova, 
I-35131 Padova, Italy}
\author{E. M. Winchester}
\author{S. J. Yennello}
\affiliation{Cyclotron Institute, Texas A\&M University, College Station, 
Texas 77843}
\author{W. Zipper}
\affiliation{Institute of Physics, University of Silesia, PL-40007, Katowice, 
Poland}
\collaboration {\bf The NIMROD collaboration}
\noaffiliation
\author{A.~Ono}
\affiliation{ Department of Physics, Tohoku University, Sendai 980-8578, Japan}

\date{\today}

\begin{abstract}
 The kinetic energy variation of emitted light clusters has been employed as 
a clock to explore the time evolution  of the temperature for thermalizing 
composite systems produced in the reactions of 26A, 35A and 47A MeV $^{64}$Zn 
with $^{58}$Ni, $^{92}$Mo and $^{197}$Au. For each system investigated, the 
double isotope ratio temperature curve exhibits a high maximum apparent 
temperature, in the range of 10-25 MeV, at high ejectile velocity. These 
maximum values increase with increasing projectile energy and decrease with 
increasing target mass. The time at which the maximum in the temperature 
curve is reached ranges from 80 to 130 fm/c after contact. For each 
different target, the subsequent cooling curves for all three projectile 
energies are quite similar. Temperatures comparable to those of limiting 
temperature systematics are reached 30 to 40 fm/c after the times 
corresponding to the maxima, at a time when AMD-V transport model 
calculations predict entry into the final evaporative or fragmentation 
stage of de-excitation of the hot composite systems. Evidence for the 
establishment of thermal and chemical equilibrium is discussed.
\end{abstract}
 


\maketitle
 
\section{Introduction}

The light particle emission which occurs during violent collisions of two 
heavy nuclei carries essential information on the early dynamics and on the 
degree of equilibration at each stage of the reaction. To obtain more 
specific information on the reaction dynamics and on the thermal evolution 
of multi-fragmenting systems produced in near Fermi energy 
collisions~\cite{suraud89,tamain96,chomaz01}, we have recently focused on 
efforts to investigate the nucleon and light cluster emission which 
occurs prior to disassembly as the system thermalizes and equilibrates. 
In some previous works, we have employed coalescence model analyses 
to probe the early dynamic evolution of the reacting 
system~\cite{cibor00,hagel00}.  In this paper, we report on the use of similar 
techniques to explore  the 
temperature evolution of hot nuclei produced in a series of reactions of   
26A, 35A and 47A MeV $^{64}$Zn projectiles with $^{58}$Ni, $^{92}$Mo 
and $^{197}$Au target nuclei using a combined 4$\pi$ 
Charged Particle -- 4$\pi$ Neutron Ball detection system. The data 
provide experimental evidence for an initial rapid thermalization 
of the incident energy, into a participant matter subsystem. The 
double isotope ratio temperature first rises to a maximum, then decreases 
as further particle emission, expansion and diffusion of the excitation 
energy into the remainder of the composite system occurs. A close 
correlation between the peak temperatures and spectral slope 
temperatures for early emitted particles is also observed, suggesting 
local chemical and thermal equilibration for this early 
emitting system. Temperatures which are comparable to those of 
limiting temperature systematics ~\cite{natowitz02_1} are reached 
about 30 to 40 fm/c after the peak temperatures are observed. 

\section{Experimental Details}

The reactions of 26A, 35A and 47A MeV $^{64}$Zn projectiles with $^{58}$Ni, 
$^{92}$Mo and $^{197}$Au target nuclei were studied at the K-500 
Super-Conducting Cyclotron at Texas A\&M University, using 
the 4$\pi$ detector array NIMROD. NIMROD consists of a 166 segment 
charged particle array set inside a neutron ball. The charged particle 
array is arranged in 12 concentric rings around the beam axis. The 
eight forward rings have the same geometrical design as the INDRA 
detector, but have less granularity ~\cite{pouthas95}. In those rings, the 
individual segments are fronted by ionization chambers (IC) filled with 
30 Torr of CF$_4$ gas. Front and back windows were made of 2.0 $\mu$m 
aluminized Mylar foil. In each of these forward rings, two of the 
segments have two Si detectors( 150 and 500 $\mu$m thick) between the IC 
and CsI detectors (super telescopes) and three have one Si 
detector(300 $\mu$m thick). Each super telescope is further divided 
into two sections. The CsI detectors are 10 cm thick Tl doped crystals 
read by photomultiplier tubes. For these detectors, a pulse shape 
discrimination method is employed to identify light 
particles~\cite{bernachi89}. In all telescopes particles are, 
identified in atomic number. In the super telescopes, all isotopes 
with atomic number Z $\leq$ 10 are clearly identified. 

The energy calibration of the Si detectors was carried out using both 
a $^{228}$Th alpha particle source and the observed punch through 
energies of identified reaction products. The punch through energies 
were calculated using a range-energy table~\cite{hubert80}. Since the 
energy losses of the lighter particles, in particular the high energy 
Hydrogen isotopes, are rather small in the Si detectors, evaluation 
of the energy deposited in the CsI crystal from the energy loss in the 
Si detectors requires special care for higher energy particles.
Therefore, an additional energy calibration was performed to measure 
energy spectra from the reaction $^{64}$Zn + $^{92}$Mo at 47A MeV.  In 
this run, Si detectors of thicknesses 1mm, backed by CsI detectors 
of three different lengths (1cm, 3cm and 5cm), were used to measure 
the inclusive energy spectra of light charged particles. The energy 
spectra were measured at all angles corresponding to those of the 12 
rings of NIMROD. The combination of thicker Si $\Delta$E detectors and 
observation of high energy punch-through points for the particles 
which traversed these thinner CsI detectors allowed us to determine 
the energy spectra with a high degree of confidence. We then used 
the $^{64}$Zn + $^{92}$Mo at 47A MeV as a standard reaction to determine 
the CsI energy calibrations for all other runs. 

Neutron multiplicity was measured with the 4$\pi$ neutron detector 
surrounding the charged particle array. This detector, a neutron 
calorimeter filled with Gadolinium doped pseudocumene, consists of 
two hemispherical end caps and a cylindrical mid-section. The mid-section 
is divided into four separate 90 degree quadrants. The hemispheres are 
150cm in diameter with beam pipe holes in the center and they are 
upstream and downstream of the charged particle array. Thermalization 
and capture of emitted neutrons in the ball leads to scintillation 
which is observed with phototubes providing event by event determinations 
of neutron multiplicity but little information on neutron energies and 
angular distributions.  Further details on the detection system, 
energy calibrations and neutron ball efficiency may be found in 
reference~\cite{wada04}.
 
During the experiment, data were taken employing two different trigger 
modes. One was a minimum bias trigger in which at least one of the CsI 
detectors detected a particle. The other was a high multiplicity trigger 
which required detected particles in 3-5 CsI detectors (depending upon 
the reaction studied).
 
\section{Data Analysis}  

An inspection of the two dimensional arrays depicting the detected 
correlation between charged particle multiplicity and neutron multiplicity 
in NIMROD (not shown), reveals a distinct correlation in which increasing 
charged particle multiplicity is associated with increasing neutron
 multiplicity.  Although there are significant fluctuations reflecting both 
the competition between different decay modes and the neutron detection 
efficiencies, these correlations provide a means for selecting the more 
violent collisions.  For the analysis reported in this paper, we have selected 
events corresponding to the largest observed neutron and charged 
particle multiplicities.  This selection corresponds to the 10\% of 
the minimum bias trigger events with the highest total multiplicity   
and emphasizes the lower impact parameter collisions. We refer to 
these events as violent collisions.  Many of the techniques applied in 
this analysis have been discussed previously in greater detail in 
references ~\cite{cibor00,hagel00,cibor01}. Only a brief summary of these 
is included in the present work. 

\subsection{Moving source analysis}

A common technique to characterize light particle emission in this energy 
range is to fit the observed spectra assuming contributions from 
three sources; a projectile-like (PLF)source, an intermediate velocity 
(IV)source and a  target-like (TLF)source.  For asymmetric collisions, 
such fits typically exhibit 
a PLF source dominance localized at high rapidity, an IV source 
dominance at mid-rapidity and TLF source emission localized at low 
rapidity~\cite{cibor00,hagel00,awes81,wada89,viola99}. In the present work,  
except for the most forward detector rings, the data are dominated by 
particles associated with the IV and TLF sources and a good reproduction of 
the observed spectra is achieved. In this analysis, the source 
velocities, temperatures, particle multiplicities and emission 
barriers for the three different sources were the parameters searched. 
 
In Figure 1, the slope temperature parameters for emission from the IV 
and TLF sources, derived from the fits, are shown for the nine reactions 
studied. For both sources the measured spectra result from a summation of 
the spectra of particles emitted over a range of time. Thus the observed 
slope temperature values are affected by the relative emission 
probabilities over that time period.  The IV source slope temperatures 
for p, d, t, $^3$He and $^4$He range from  T $\sim 7$ to 17 MeV for the 

different systems studied. The temperatures from the particles with 
A $\leq$ 3 are quite similar. They follow the trends of earlier reported 
values for pre-equilibrium emission at such projectile 
energies~\cite{awes81,wada89,fox88}.  The IV source temperatures derived 
from the alpha spectra are typically lower than those measured for the other 
particles.  These softer slopes for the alpha particles appear to reflect 
larger relative contribution of lower energy alphas which are attributed to 
the IV source in the fitting procedure.  The slope parameters for the TLF 
sources are much lower, in the range of T$\sim2-6$ MeV.  For this source, the 
apparent temperatures for alpha emission are the highest. Such an effect 
has previously been noted and attributed to the relatively higher 
emission probabilities for alpha particles in the early stage of the 
evaporation cascade ~\cite{hagel88,gonin90}.

\begin{figure}
\includegraphics[scale=0.4]{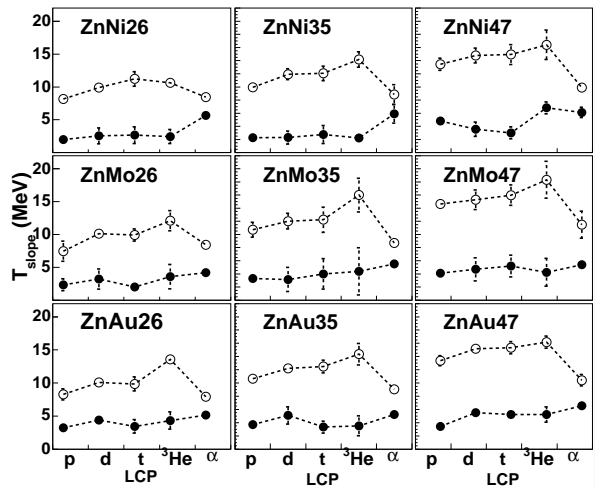}
%\vspace{-2.0truecm}
\caption{\footnotesize Slope temperatures from 3-source fits to the 
experimental spectra.  Open circles represent apparent temperatures for 
the emission from the IV source. Filled circles represent apparent 
temperatures for the emission from the TLF source.}
\label{slopeTemperatures}
\end{figure}

\subsection{Temperature Determinations}
Given the continuous dynamic evolution of the system, source fit parameters 
should be considered as providing only a schematic picture of the emission 
process.  Nevertheless, the information derived can be very instructive. We 
have employed such analyses to estimate the multiplicities and energy 
removed at various stages of the reaction. To follow the time evolution 
of the system in more detail, a more sophisticated analysis of the particle 
emission is necessary. Both theoretical models ~\cite{lukasik93,ono99} 
and experiments ~\cite{he98,gelderlos95} indicate that the early collision 
dynamics leads to a correlation between emission time and energy for the 
early emitted particles. This correlation can be exploited to follow the 
time evolution of the system. We have previously used this correlation in 
coalescence model studies of several systems~\cite{cibor00,hagel00,cibor01}. 
For the present systems, extensive calculations have been made employing 
the AMD-V model of Ono~\cite{ono99}.  Many of the results of the 
AMD-V calculations for the present systems are compared to experimental 
observables in reference ~\cite{wada04}. In the present work, we have 
employed the calculated correlation between particle energy and time 
predicted by those AMD-V calculations to calibrate the emission time 
scales for these reaction systems. 

 At intermediate energies, the observed spectral slope parameters derived 
from the source fits are not adequate as measures of the temperature 
evolution, as the observed spectra are convolutions of the spectra at 
different emission times and excitation energies and include high energy 
particles which are emitted prior to the achievement of thermal equilibrium.  
For a system at chemical and thermal equilibrium at a suitably low density, 
Albergo {\it et al.}~\cite{albergo85} have shown that the temperature of the 
emitting system can be derived directly from the first chance emission 

double isotope yield ratios of two adjacent isotopes of two different 
elements.  In a more recent work by Kolomiets {\it et al.}~\cite{kolomiets97}, 
essentially the same result is derived when only thermal equilibrium is 
initially assumed.  Therefore, to characterize the temperature at a particular 
emission time, we have employed double isotope yield ratio measurements. 
In the case of strong system evolution, double isotope yield ratio 
temperatures derived from integrated yields are certainly suspect if 
the isotopes being utilized are in fact produced at very different 
times or by different mechanisms.  However, if chemical equilibrium 
is achieved and the particles corresponding to particular emission 
times can be selected, derivations of double isotope yield ratio 
temperatures as a function of emission time should allow us to follow 
the temperature evolution of the system. 

To focus on the early evolution of the temperature, we have first selected 
clusters observed at mid-rapidity, i.e., specifically those detected at 
angles between 70$^o$ and 80$^o$ in the IV source frame. In this way, we 
attempt to isolate the emission associated with the IV source which 
occurs during the thermalization stage of the 
reaction~\cite{cibor00,hagel00,cibor01}. We have then made double isotope 
ratio temperature determinations as a function of ejectile velocity 
in the IV frame. The velocities used are the ``surface velocities'' of 
the emitted particles. The surface velocity, $V_{surf}$, is defined as the 
velocity of an emitted species at the nuclear surface, prior to 
acceleration in the Coulomb field ~\cite{awes81,hagel00}. $V_{surf}$, is 
obtained in our analysis by subtraction of the Coulomb barrier energy 
derived from the source fits. Since the early emitted light particle 
energies are strongly  correlated with emission times, and evaporative 
or secondary emission contribute to the spectra primarily at the lower 
kinetic energies, the yields of higher energy particles should be 
relatively uncontaminated by later emission processes. This is an 
important advantage in such double isotope yield ratio determinations. 
At mid-rapidity there is little contribution from the PLF source. 
There is, however some observed contribution from the TLF source at 
low velocities in the IV source frame. Since the three source fits are 
only approximations to the emission from the continuously evolving 
system, particles in this low velocity range may be viewed alternatively 
as the last particles emitted from the IV source or the earliest from 
the TLF source. In the following analysis yields assigned to the TLF 
source have been subtracted from the experimental yields.

The temperatures employed are T$_{HHe}$, derived from the yields of d, t, 
$^3$He and $^4$He clusters. For particles emitted from a single source of 
temperature, T, and having a volume Maxwellian spectrum 
($\sqrt{\epsilon}\exp{-\epsilon/T}$), 
where $\epsilon$ is the particle energy, the HHe double isotope yield ratio 
evaluated for particles of equal $V_{surf}$, is $\sqrt{\frac{8}{9}}$ 
times the ratio derived from either the integrated particle yields or 
the yields at a given energy above the barrier ~\cite{albergo85}. Thus
\begin{eqnarray}
T=\frac{14.3}{\ln{\Big(\sqrt{\frac{9}{8}}\big(1.59R_{v_{surf}}\big)}\Big)}
\end{eqnarray}
where the constants 14.3 and 1.59 reflect binding energy, spin, masses and 
mass differences of the ejectiles. If Y represents a cluster yield, 
$R(V_{surf}) = Y_dY_{^4He}/ Y_tY_{^3He}$  for clusters with the same 
surface velocity.  

\subsection{Calibration of Timescales}
To calibrate the time-scale associated with our data,we have employed 
results of the AMD-V calculations ~\cite{wada04}. In the AMD-V calculations, 
the particle emission starts at times near 50 fm/c, and varies depending on 
projectile velocity and entrance channel masses. From that point, the 
calculations predict an initial rapid decrease of the average kinetic 
energies of the emitted particles with increasing time of emission. This 
is followed by a much slower rate of decrease at later times, 140 to 
200 fm/c for the systems studied. The time of the transition from rapid 
to relatively slow kinetic energy depends on projectile velocity and 
entrance channel mass. Such trends are typically observed in transport 
model calculations. For the four systems studied, we have derived from 
AMD-V calculations, which covered the range of time from the time of 
contact up to to 300 fm/c, the correlation between average emission 
times of emitted neutrons and protons and their energies. For each 
reaction the resultant time-surface velocity relationship is represented 
in Figure 2. As is seen there, the calculations indicate a near linear 
decrease of V$_{surf}$ , from near  projectile velocity into the 3 to 
3.5 cm/ns range as the average emission times increase from $\sim 50$ fm/c 
to 150 fm/c. In the following we have employed these correlations to determine 
the average emission times corresponding to particular observed values of 
V$_{surf}$ in the experiment.  However, since below 3.5 cm/ns the experimental 
data contain large contributions from TLF evaporative emission, the 
sensitivity of the emission energy to time is significantly reduced, 
and we do not attempt to assign emission times for particles with surface 
velocities below 3.5 cm/ns. 

\begin{figure}
\includegraphics[scale=0.4]{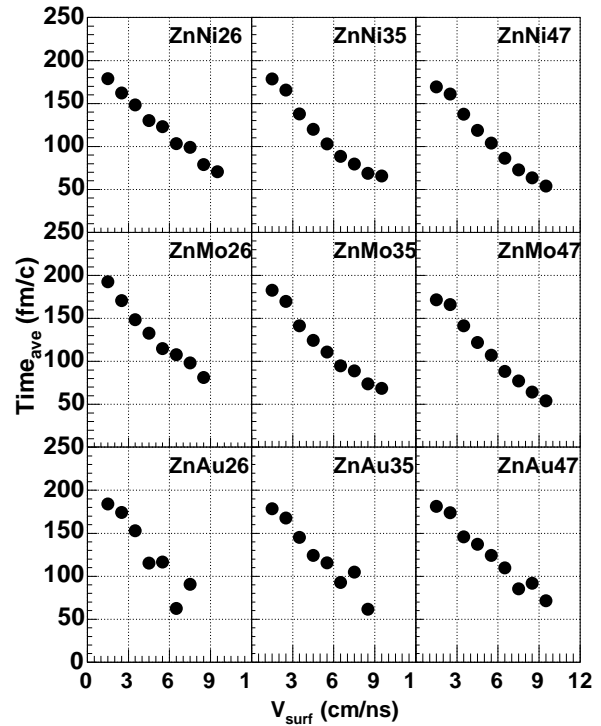}
%\vspace{5.0truecm}
\caption{\footnotesize Correlation of average emission time with surface 
velocity for early emitted nucleons as calculated by the AMD-V code. Solid 
symbols depict the results for the nine different reactions as labeled.
}
\label{timecal}
\end{figure}

\section{Results and Discussion}

 We present, in Figure 3, experimental results for the double isotope ratio 
temperatures as a function of velocity in the IV frame. For each system 
investigated, the double isotope ratio temperature determination exhibits 
a high maximum temperature in the range of 10-27 MeV. These maximum 
temperatures are much higher than the limiting temperatures determined 
from caloric curve measurements in similar reactions~\cite{natowitz02_1}.  
  In each
case, the apparent temperature 
decreases monotonically on either side of this maximum.  The AMD-V model 
calculations~\cite{ono99,wada04} indicate a significant slowing  in the 
rate of kinetic energy change in the 3-3.5 cm/ns velocity range. The 
shaded vertical bars in Figure 3 indicate that velocity value.  These 
points signal the end of the IV (or pre- equilibrium) emission stages. 
At lower velocities, the slower nuclear de-excitation modes evaporation, 
fission and/or fragmentation determine the properties of the ejectile 
spectra.  At these lower velocities, values of T$_{HHe}$ are 3 to 4 MeV,
similar to those spectral integrated values seen in other 
experiments~\cite{kolomiets97,xi98_1,xi98_2,viola04,wada97}. These values 
are also very similar to T$_{HHe}$ temperatures calculated when the sequential 
evaporation code GEMINI~\cite{charity88} is used to simulate the 
de-excitation of the TLF source~\cite{hagel88}. We take this as further 
evidence that the spectra at these lower velocities still contain 
contributions from late stage evaporation. Temperatures derived from 
the yield ratios in this velocity range require corrections for 
secondary decay effects.
        
\begin{figure}[b]
\includegraphics[scale=0.45]{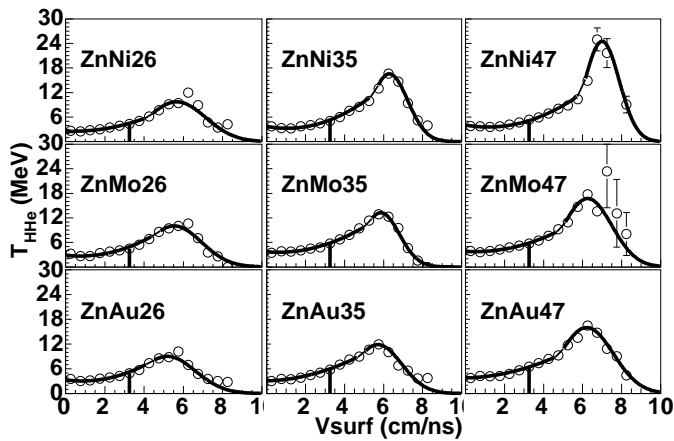}
%\vspace{-7.0truecm}
\caption{\footnotesize T$_{HHe}$ vs surface velocity. See text. Horizontal 
bars are at 3-3.5 cm/ns, taken to be limit for time derivations. Solid lines 
indicate fits to data.}
\label{albergoTemp}
\end{figure}
 
Figure 4 presents the derived T$_{HHe}$ temperatures as a function of time. 
While all nine reactions show a qualitatively similar evolution with time, we 
now see that, for each energy, the time at which the maximum in the 
temperature curve is reached increases with increasing target mass and 
decreases with increasing projectile energy. The former observation suggests 
a longer period required for establishment of a thermal and/or chemical 
equilibrium as the total system size increases while the latter observation 
suggests a more rapid thermalization of the initial projectile energy for the 
initially faster projectiles. The dynamic transport calculations indicate 
that the condition of thermal equilibrium of the whole system is not yet 
established as the earliest ejectiles are emitted. Non equilibrium effects 
may be most evident in the particularly high apparent peak temperatures 
for the 35A and 47A MeV $^{64}$Zn + $^{58}$Ni case. The AMD-V calculations for 
the different systems predict a higher degree of transparency in those 
reactions\cite{wada04}.
                                    
\begin{figure}
\includegraphics[scale=0.4]{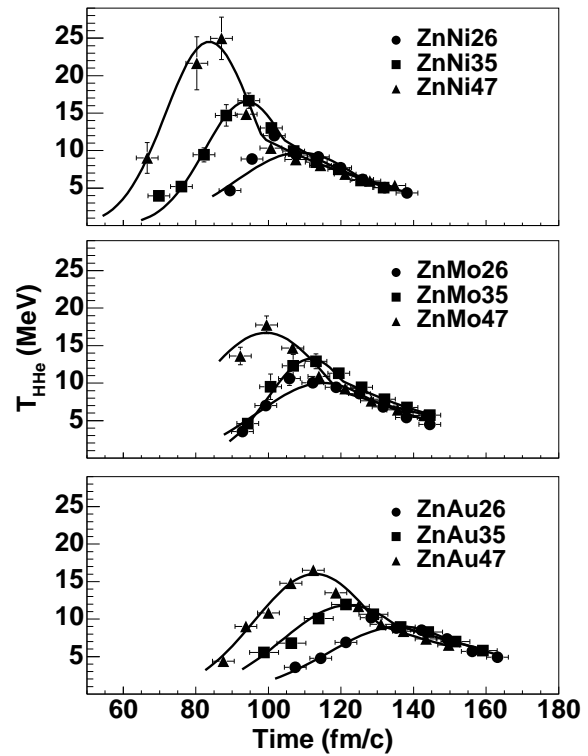}
%\vspace{-5.0truecm}
\caption{\footnotesize T$_{HHe}$ vs time. See text. Times 
terminate at points corresponding to surface velocities of 3-3.5 cm/ns.}
\label{albergoTempVsTime}
\end{figure}

 Figure 4 also indicates that the time of entry into the evaporation or 
disassembly stage, taken to be the time corresponding to surface velocities 
of 3.5 cm/ns, increases with target mass, from $\sim135$ fm/c for the Ni 
target to $\sim165$ fm/c for the Au target. We note that at such times the 
temperatures are very similar to the limiting temperatures derived from a 
systematic investigation of caloric curve measurements ~\cite{natowitz02_1}.  
Except for the $^{64}$Zn + $^{58}$Ni reactions at 35A and 47A MeV, the time 
for the initial cooling stage, i.e., the time difference between that 
corresponding to the maximum in the temperature and that corresponding to the 
start of the evaporation stage, is 30 to 40 fm/c. 

\subsection{Interpretation of Temperature Evolution Curves}

The experimental temperature curves show an initial rise to a maximum and a 
subsequent decline. It is tempting to interpret the initial rise as 
reflecting the early rate of conversion of projectile kinetic energy 
into thermal energy of the composite system. However, it appears more likely 
that the double isotope temperature is not accurately reflecting the 
temperature of the emitting system at earlier times.  
The time required for the establishment of chemical equilibrium %which 
is presumably longer than that required for thermalization.  We are not 
able to separate these with the present data. Rather, in this section 
we address the interpretation of the temperatures and the degree to 
which they can be taken as reflecting thermal and chemical equilibration, 
at least locally if not globally.

\subsection{Comparison of T$_{HHe}$ with T$_{slope}$ from source fits}

If both thermal and chemical equilibrium are achieved and the density is 
not too high~\cite{majka97}, an agreement between the thermal temperature 
and the double isotope ratio temperature T$_{HHe}$ can be expected. 

For the TLF source the fit slope parameters, T$_{slope}$  
presented in Figure 1 are normally lower than the latest time 
T$_{HHe}$ temperatures seen in Figure 4. This is not surprising since 
the source fit can be expected to return only an apparent temperature 
reflecting the entire cooling stage of the TLF source.  In our previous 
work~\cite{hagel88,gonin90,wada89}, we have found that the slope parameters 
for the alpha particle emission from the TLF source most closely 
approximates the intial thermal temperature of this source, reflecting 
the higher fraction of the alpha emission in the earlier part of the 
de-excitation cascade. The alpha particle slope parameters are presented 
in column 5 of Table I. The relationship between these thermal fit 
apparent temperatures and the late time chemical temperatures obtained 
from the TLF isotope ratio temperatures, as presented in column 6, is quite 
reasonable, given the associated uncertainties.

\begin{table*}
\caption{Temperatures and Excitation Energies}
\begin{tabular}{|c|c|c|c|c|c|cc|}
\hline
Target & Projectile & T$_{slope}(Avg)$ & Max T$_{HHe}$ & 
T$_{slope}$(TLF) & T$_{HHe}$(TLF) & E$^*_{max}$ & T$_{max}$ \\
&Energy & IV  &     & alpha & ($v_{surf}=3.5$cm/ns)  &  &A$_{part}$=A$_{tot}$\\
   &   &  &   &  &  &  &  $\rho=\rho_0$  \\
%\hline
& (MeV/u) & (MeV) & (MeV) & (MeV) & (MeV) & (MeV/u) & (MeV) \\
\hline
$^{58}$Ni & 26.0 & $9.68\pm0.68$ & $9.74\pm0.25$ & $5.66\pm 0.30$ & $4.34\pm 0.40$ & 5.92 & 9.49 \\
          & 35.0 & $11.4\pm1.05$ & $16.6\pm0.99$ & $5.91\pm 1.43$ & $5.06\pm 0.45$ & 8.17 & 11.1 \\
          & 47.0 & $13.7\pm 1.28$& $24.8\pm2.57$ & $6.12\pm 0.80$ & $5.34\pm 0.46$ & 11.2 & 13.0 \\
\hline
$^{92}$Mo & 26.0 & $9.60\pm 0.97$& $10.0\pm0.42$ & $4.18\pm 0.50$ & $4.50\pm 0.40$ & 5.55 & 9.19 \\
          & 35.0 & $11.9\pm 1.44$& $13.3\pm0.99$ & $5.50\pm 0.50$ & $5.74\pm 0.54$ & 7.73 & 10.8 \\
          & 47.0 & $15.1\pm 1.73$& $17.6\pm1.13$ & $5.41\pm 0.30$ & $5.71\pm 0.53$ & 10.6 & 12.7 \\
\hline
$^{197}$Au& 26.0 & $9.95\pm 0.69$& $9.02\pm0.18$ & $5.13\pm 0.13$ & $4.92\pm 0.45$ & 3.96 & 7.76 \\
          & 35.0 & $11.8\pm 0.81$& $11.9\pm0.31$ & $5.22\pm 0.35$ & $5.81\pm 0.54$ & 5.62 & 9.25 \\
          & 47.0 & $14.1\pm 0.78$& $15.7\pm0.54$ & $6.54\pm 0.32$ & $6.53\pm 0.60$ & 7.84 & 10.9 \\

\hline
\end{tabular}
\end{table*}

The observed time evolution of T$_{HHe}$ also prompts us to inquire 
about the relationship between the maximum temperatures and the slope 
temperature parameters which characterize the ejectile kinetic energy 
spectra of the IV sources. These latter are also determined from the 
three source fits to the experimental spectra. As is seen in Table I 
and Figure 5, there is actually a close agreement between these two 
values for all but the 35 and 47A MeV $^{64}$Zn + $^{58}$Ni reactions. The 
apparently high values of the peak temperatures for these last two 
reactions were already seen above. As noted there, this may 
reflect non-equilibrium effects resulting from the very high degree 
of transparency for those two reactions.

At first glance, the good agreement seen for the rest of the 
reactions still seems surprising because the IV spectrum is known to result 
from a convolution of the ejectile emission from the evolving system. 
Thus, it depends on the time-dependent rates of both the ejectile 
emission and the temperature evolution of the system. However, if 
the requirement that the global fit to the IV source reproduce the 
high energy tail of the IV spectrum plays a dominant role in determining 
the overall slope temperature parameter of the fit, this result is 
understandable. In such a case, the agreement seen between 
this thermal slope  parameter and the chemical temperature determined 
from the isotope yield ratios for the lowest energy 26A MeV 
$^{64}$Zn + $^{58}$Ni reaction and for the $^{64}$Zn + $^{92}$Mo and 
$^{64}$Zn + $^{197}$Au reactions at all three projectile energies 
provides a strong indication that a simultaneous thermal and chemical 
equilibrium is achieved in the emitting system, at least locally, at 
times corresponding to those of the peak temperatures. 

\begin{figure}
\includegraphics[scale=0.4]{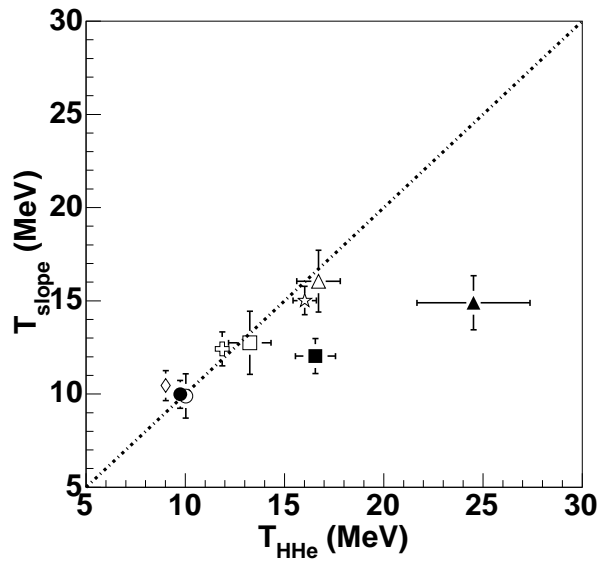}
%\vspace{-2.0truecm}
\caption{\footnotesize Relationship between T$_{HHe}$ peak values and 
T$_{slope}$ of the IV source. Dotted line represents locus of equality 
of the two temperatures.  Open symbols show results for the reactions with 
Mo and Au targets.  Solid symbols show results for reactions with Ni tragets}
\label{TPeakVsTSlope}
\end{figure}

\subsection{Model estimates of peak temperature and source sizes}

We have attempted to make a first order estimate of the maximum 
temperature to be expected. 
In the AMD calculations for our systems, only a small fraction of the 
available mass and excitation energy has been removed from the system by 
pre-equilibrium emission at the times which we associate with the observed 
maximum temperatures~\cite{wada04}. For the reaction being considered, 
we then ignore the mass loss and estimate the maximum thermal excitation 
energy to be E$^*$ = E$_{CM}$ + Q, where E$_{CM}$ is the available 
Center of Mass energy and Q, the fusion reaction Q value. Excitation 
energies per nucleon of 5.9 to 11.2 MeV/u are obtained in this way.  The 
results of these calculations are presented in column 7 of Table I.  At the 
times corresponding to the maximum observed temperature, the AMD-V 
calculations show the 
composite systems to have rebounded from an initial small compression and 
the system density, $\rho$, to be at or below normal. If $\rho/\rho_0$ = 1, we 
can estimate the maximum temperature which we might expect. Here we 
assume a uniform normal density Fermi gas~\cite{preston62} with a 
Fermi Energy which we take from the interpolation or extrapolation of 
the values reported in reference ~\cite{moniz71}. Since the system is 
quite excited, we further assume a nucleon effective mass of 
1~\cite{shlomo91}. Ignoring the small mass loss expected, we calculate 
the temperature, T, from E$^* = aT^2$, where $a$, the level density 
parameter, is determined from the Fermi energy.   Here 
$a$ = A/15.2~MeV$^{-1}$ is 
used~\cite{moniz71}.  The temperatures thus derived are presented in 
Table I, column 8.  While the calculated temperatures reported in 
column 8 are indeed significantly higher than limiting termperatures of 
caloric curve measurements, they are not as high as the observed maximum 
temperatures in column 4. 

The calculated values in column 8 follow from an assumed thermalization of 
the entire system. In a recent investigation, Sood and Puri have employed a 
QMD transport model to calculate the maximum and average temperatures and 
densities achieved in symmetric or near symmetric heavy ion collisions at 
E$_{bal}$, the balance energy corresponding to the transition from positive to 
negative flow~\cite{sood03}.  Their calculation of the maximum temperature, 
based upon a local density approximation for the matter contained in a sphere 
of 2 fm radius around the center of mass of the system, clearly indicates 
that the entire system is not equilibrated at the early times.  In general, 
the available center of mass energies in the calculations of 
reference~\cite{sood03} are somewhat higher than those of our reactions with 
47A MeV $^{64}$Zn and the calculations are made for 
varying impact parameters. Nevertheless, the maximum double isotope 
ratio temperatures derived in the present experimental study are quite 
comparable to those reported in reference \onlinecite{sood03}. This is 
particularly noteworthy because the calculations by Sood and Puri 
strongly suggest the presence of an initial hot, locally equilibrated, 
participant zone surrounded by colder spectator matter and the derived 
temperatures  are interpreted as representing 
a local thermal equilibrium. A similar picture is obtained in the AMD-V 
calculations of reference~\cite{wada04}. Further, such a result is 
consistent with results of  earlier experimental studies of pre-equilibrium 
emission which found that the IV spectra could be equally well modeled 
either within the framework of nucleon-nucleon collision dynamics or as 
emission from a hot thermalized participant zone~\cite{awes81,fox88}. 

If, in fact, the early system consists of both participant (nascent 
fireball ~\cite{mekjian78}) and spectator matter then, initially, the 
available energy may be distributed only over a subset of the nucleons. 
Further, the density for this subsystem need not be $\rho_0$. Assuming 
still that such a hot participant zone may be modeled as a uniform 
density Fermi gas allows us to write the more general 
expression~\cite{natowitz02_2,viola04}
$$T=\sqrt{(K_0(\rho/\rho_0)^{\frac{2}{3}} (E_x/A_{part})},$$

where $K_0$ is the inverse level density parameter at normal density (in an 
excited nucleus) and A$_{part}$ is the mass number of the participant zone. 
Taking $K_0$ to be 15.2 MeV and the observed peak temperatures from column 4 
of table I allows us to calculate $A_{part}$ as a function of $\rho/\rho_0$ 
for each system studied at each projectile energy. The results are presented 
in Figure 6 (a)-(c), where the calculated $A_{part}$, normalized to the 
projectile mass, is plotted against $\rho/\rho_0$. Some initial compression 
and subsequent expansion is predicted by the AMD-V calculation.  The density 
at the time of the experimental peaking of the temperature is expected to
be less than normal density.  The results of this calculation indicate that 
the number of participant nucleons at the time of the peaking of the 
temperature decreases with increasing projectile energy and increases with 
increasing target mass.  The values of the ratios which would result from 
participation of the total entrance channel masses are indicated by the 
horizontal lines in the figure.  Most experimental values are well below 
those. These results suggest the existence of a hot zone at early times. 
For the case of 47A $^{64}$Zn + $^{197}$Au, results of the AMD-V calculations 
reported in reference\cite{wada04} indicate the existence of such a zone at 
early time, with  a mass number about twice that of the projectile. As seen in 
Figure 6, the Fermi gas estimate for this system at below normal densities 
is quite close to that. 

\begin{figure}
\includegraphics[scale=0.4]{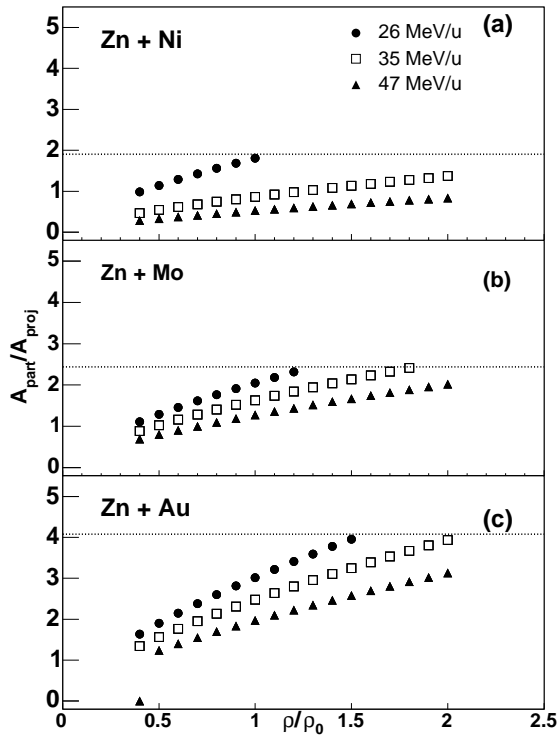}
%\vspace{5.0truecm}
\caption{\footnotesize Calculated values for the correlation of 
A$_{part}$/A$_{proj}$ and $\rho/\rho_0$ consistent with measured peak
temperatures.
}
\label{massEvol}
\end{figure}

While the possibility of emission from a thermalized hot zone is one of 
the possible interpretations which has previously been suggested in earlier 
comparisons of dynamic and thermal pictures of pre-equilibrium emission 
in similar collisions~\cite{awes81,fox88}, in the present study it is 
inferred not from the slope parameters of the pre-equilibrium source but 
rather from the peak value of the temperature, T$_{HHe}$. 
The double isotope ratio temperature measurements assume chemical 
equilibrium.  We return to this point in the following section. 

\section{Summary and Conclusions}

The kinetic energy variation of emitted light clusters has been employed 
as a clock to explore the time dependence of the temperature evolution of 
thermalizing composite systems produced in the reactions of 26A, 35A and 
47A MeV $^{64}$Zn with $^{58}$Ni, $^{92}$Mo and $^{197}$Au. For each 
system investigated, the double isotope ratio temperature curve exhibits 
a high maximum apparent temperature, in the range of 10-25 MeV. The 
maximum values increase with increasing projectile energy and decrease 
with increasing target mass. These maxima occur at times from 80 to 
130 fm/c after the nuclei contact. They are much higher than the 
limiting temperatures determined from caloric curve measurements in 
similar reactions~\cite{natowitz02_1}. For most of the reactions 
studied, a close correlation is observed between the peak temperatures 
for early emitted particles, obtained from double isotope ratios, and 
the spectral slope temperatures for the pre-equilibrium (IV) source. 
The data indicate that at least a local thermal and chemical equilibrium 
is established during these times. After peaking, the temperatures 
decrease rapidly, apparently reflecting particle emission, diffusion 
of the excitation energy into the remaining system and expansion. For 
each individual target nucleus, the later portions of the cooling curves 
for all three projectile energies are very similar, indicating that hot 
nuclei with similar properties are produced. Temperatures comparable to 
those derived from limiting temperature systematics are reached 30 to 
40 fm/c after the times corresponding to the maxima, at the times when AMD-V 
transport model calculations predict entry into the final evaporative 
or fragmentation decay of the hot composite system. 

As a final comment, we note that the present data suggest that if the 
Z=1 and Z=2 ejected light particles are taken to represent the gaseous 
phase, as is usually assumed, the reaction dynamics of central collisions 
may itself lead to a natural situation in which the gas is not outside 
the liquid matter but initially confined inside the liquid matter, 
perhaps facilitating the establishment of a liquid-gas equilibrium. 
This could somewhat mitigate arguments that the concept of establishment 
of a liquid gas equilibrium is not tenable in a nuclear collision as 
there is no container to constrain the gas. Of course, more recently 
it has also been argued, on the basis of detailed balance, that the 
actual physical equilibrium is not necessary ~\cite{moretto02}. 

\section{Acknowledgements}

This work was supported by the United States Department of Energy under Grant 
\# DE-FG03- 93ER40773 and by The Robert A. Welch Foundation 
under Grant \# A330.  The work of JSW is also partially supported by the 
NSFC under Grant \# 10105011.

{}

\end{document}